%% file: mainCA.tex
\def\lncs{1}
\ifnum\lncs=0
	\documentclass[11pt, notitlepage]{article}
	\usepackage[a4paper, margin=2.54cm, includefoot]{geometry}
\else
	\documentclass{llncs}
\fi

\usepackage{PVC}
\usepackage{james}

\ifnum\lncs=0
	\title{Publicly Verifiable Outsourced Computation with a Trusted Third Party}
	\author{James Alderman\thanks{James.Alderman.2011@live.rhul.ac.uk} \and Carlos Cid\thanks{Carlos.Cid@rhul.ac.uk}
	 \and Jason Crampton\thanks{Jason.Crampton@rhul.ac.uk} \and Christian Janson\thanks{Christian.Janson.2012@live.rhul.ac.uk} \\
  \multicolumn{1}{p{\textwidth}}{\centering{Information Security Group, Royal Holloway, University of London}}}
\else
	\title{Publicly Verifiable Outsourced Computation with a Key Distribution Centre}
	\author{James Alderman\thanks{James.Alderman.2011@live.rhul.ac.uk} \and Carlos Cid\thanks{Carlos.Cid@rhul.ac.uk}
	 \and Jason Crampton\thanks{Jason.Crampton@rhul.ac.uk} \and Christian Janson\thanks{Christian.Janson.2012@live.rhul.ac.uk}}
	\institute{Information Security Group, Royal Holloway, University of London}

\fi
\date{}

\providecommand{\keywords}[1]{\textbf{\textit{Keywords---}} #1}

\begin{document}

\setlist[itemize]{noitemsep, topsep=0pt}
\setlist[enumerate]{noitemsep, topsep=0pt}

\maketitle

\input{Abstract}

\ifnum\lncs=0 %
\input{Introduction}

	\input{BackgroundVC}
\input{SCdefs}

\input{Construction}
	\input{Conclusion}
	\section*{}
	\bibliographystyle{abbrv}
	\bibliography{index.bib,jamesindex.bib}
	\appendix
	\input{Appendix}
\else %

\input{Introduction}\input{BackgroundVC}\input{SCdefs}
	\input{Theorem}
	\section*{}
	\bibliographystyle{abbrv}
	\bibliography{index.bib,jamesindex.bib}
	\appendix

\input{Construction}\input{AppendixGames}

\fi

\end{document}

%% file: Abstract.tex
\begin{abstract}

The combination of software-as-a-service and the increasing use of mobile devices gives rise to a considerable difference in computational power between servers and clients. Thus, there is a desire for clients to outsource the evaluation of complex functions to a server and to be able to verify that the resulting value is correct. Previous work in this area of Publicly Verifiable Outsourced Computation (PVC) requires a costly
pre-processing stage. However, in many practical situations multiple clients will be interested in the same set of core functions and will make use of the same servers. Thus, the pre-processing phase may be performed many more times than is necessary. In this paper we introduce a Key Distribution Center (KDC) that handles the generation and distribution of the keys that are required to support PVC, thereby eliminating this redundancy. We define a number of new security models and functionalities that arise with the introduction of the KDC,
and present a construction of such a scheme built upon Key-Policy Attribute-based Encryption.

\end{abstract}
\ifnum\lncs=1
\keywords{Publicly Verifiable Outsourced Computation,
Key Distribution Center,
Key-policy Attribute-based Encryption,
Revocation}
\fi

%% file: Introduction.tex
\section{Introduction}
\label{sect:intro}

\ifnum\lncs>0 %

It is increasingly common for mobile devices to be used as general computing devices.
There is also an increasing trend towards cloud computing and enormous volumes of data (``big data'') which mean that computations may require considerable computing resources.
In short, there is, increasingly, a discrepancy between the computing resources of end-user devices and the resources required to perform complex computations on large datasets.
This discrepancy, coupled with the increasing use of software-as-a-service, means there is a requirement for a client device to be able to delegate a computation to a server.

Consider, for example, a company that operates a ``bring your own device'' policy, enabling employees to use smartphones and tablets.
It may not be possible for these devices to perform complex computations locally.
Instead, a computation is outsourced over some network to a more powerful server (possibly outside the company, offering software-as-a-service, and hence untrusted) and the result of the computation is returned to the client device.
Another example scenario arises in the context of battlefield communications where each member of a squadron of soldiers is deployed with a reasonably light-weight computing device.
The soldiers gather data from their surroundings and send it to regional servers for analysis before receiving  tactical commands based on results.
Thus a soldier must have an assurance that the command has been computed correctly and by a trusted party.
A final example could consider sensor networks where lightweight sensors transmit readings to a more powerful base station to compute statistics that can be verified by an experimenter. 

In simple terms, given a function $F$ to be computed by a server $S$, the client sends input $x$ to $S$, who should return $F(x)$ to the client.
However, there may be an incentive for the server (or an imposter)
to cheat and return an invalid result $y \ne F(x)$ to the client.  The acceptance of an incorrect result
may have an advantage for the server, or the server may be too busy or may not wish to devote resources to perform the computation.
Thus, the client wishes to have some assurance that the result $y$ returned by the server is, in fact, $F(x)$.
This problem, known as \emph{Verifiable Outsourced Computation} (VC), has attracted a lot of
attention in the community recently (see Sect.~\ref{sect:related} for a brief overview). 
Many current schemes have an expensive pre-processing stage run
by the client, which should be amortised over many function evaluations over distinct inputs.
Note, however, it is likely that many different clients will be interested in outsourcing computations, and also that the functions of interest to each of the clients will substantially overlap, as in the ``bring your own device'' scenario discussed above.
It is also conceivable that the number of computation servers offering to perform such computations will be relatively low (limited to a reasonably small number of trusted companies with plentiful resources).
Thus, it is easy to envisage a situation in which many computationally limited clients wish to outsource the computation of the same function $F$ to the same server, yet each
must individually expend considerable resources to run the setup phase.

Our main contribution in this paper is to introduce a \ca\ (\caa), that is responsible for running the setup stage on behalf of \emph{all} clients.
Thus, the expensive algorithm is executed just once and by the more capable \caa, rather than multiple times by restricted client devices.
We consider two example settings: one is a straightforward generalisation of the previously considered model where clients send computations directly to an available server; in the second setting, we allow a pool of computational servers governed by some managing entity. Clients submit jobs to this pool and the manager distributes work according to a scheduling policy or a bidding process, and the result is returned to the client -- thus the client may not require knowledge of the server identity or credentials beforehand.

We give definitions for a new framework of \PVC\ that both removes redundancy and facilitates additional functionality (such as revoking misbehaving servers), including several new security notions.
We also give a provably secure instantiation that meets the new definitions.
In the manager model, we allow for ``blind verification'' by the manager or other entities, a form of output privacy, such that he learns whether the result is valid but not the value of the output. Thus he may reward or punish servers appropriately without learning function outputs.

It may be tempting to suggest that the \caa, as a trusted entity, performs all computations itself. 
However we believe that this is not a practical solution in many real world scenarios, \eg\ the \caa\ could be an authority within the organisation responsible for user authorisation that wishes to enable workers to securely utilise cloud-based software-as-a-service.
As an entity within the boundaries of the organisation, performing all computations would negate the benefits gained from outsourcing computations to externally available powerful servers.
Additionally, as an authority on users and keys, the \caa\ may have simultaneous responsibilities in other systems, and we minimise its workload to key generation and revocation only.

\else

It is increasingly common for mobile devices to be used as general computing devices.
There is also an increasing trend towards cloud computing and enormous volumes of data (``big data'') which mean that computations may require considerable computing resources.
In short, there is, increasingly, a discrepancy between the computing resources of end-user devices and the resources required to perform complex computations on large datasets or to evaluate complex functions.
This discrepancy, coupled with the increasing use of software-as-a-service, means there is a requirement for a client device to be able to delegate a computation to a server.

Consider, for example, a company whose database contains many terabytes of
data and operates a ``bring your own device'' policy, enabling employees to use smartphones and tablets.
It may not be possible for these devices to perform complex computations locally.
Instead, a computation is outsourced over some network to a more powerful server (possibly outside the company, offering software-as-a-service and hence untrusted) and the result of the computation is returned to the client device.

Another example scenario arises in the context of battlefield communications where each member of a squadron of soldiers is deployed with a reasonably light-weight computing device and a communications link to headquarters.
The soldiers gather data from their surroundings and send it back to headquarters for analysis before receiving  tactical commands based on results computed from the environmental data.
Thus a soldier must have an assurance that the command has been computed correctly and by a trusted party.

In simple terms, given a function $F$ to be computed by a server $S$, the client sends input $x$ to $S$, who should return $F(x)$ to the client.
However, there may be an incentive for the server (or an imposter)
to cheat and return an invalid result $y \ne F(x)$ to the client.  The
server could, for example, be interested in convincing the client to accept an incorrect result
because this could have an advantage for the server, or the server may be too busy or may not wish to devote the computational resources to perform the computation.
Thus, the client wishes to have some assurance that the result $y$ returned by the server is, in fact, $F(x)$.

This problem, known as \emph{Verifiable Outsourced Computation} (VC), has attracted a lot of
attention in the community recently (see Sect.~\ref{sect:related} for a brief overview). 
Many of the current schemes have an expensive pre-processing stage run
by the client.
The idea is that the client will submit multiple (distinct) inputs to the server for evaluation; thus an expensive (but one-time) setup stage per
function is acceptable, as the cost will be amortized over subsequent function evaluations.
Note, however, it is also likely that many different clients will be interested in outsourcing computations, and also that the functions of interest to each of the clients will substantially overlap, as in the ``bring your own device'' scenario discussed above.
It is also conceivable that the number of available computation servers offering to perform such computations will be relatively low (limited to a reasonably small number of trusted companies with plentiful resources).
Thus, it is easy to envisage a situation in which many computationally limited clients wish to outsource the computation of the same function $F$ to the same server, yet each
must individually expend considerable resources to run the setup phase.

Our first contribution in this paper is to introduce a Key Distribution Center (\caa), who is responsible for running the setup stage on behalf of \emph{all} clients.
Thus, the expensive algorithm is executed once and by the more capable \caa, rather than multiple times by restricted client devices.
We then explain how an existing VC scheme can be extended to provide an instantiation of the new approach.
Finally, we introduce several new security notions appropriate for the modified protocol and prove that our scheme is secure with respect to these notions.
\fi

%% file: BackgroundVC.tex
\section{Verifiable Computation Schemes and Related Work}
\label{sect:backgroundvc}

The concept of non-interactive verifiable computation was introduced by Gennaro \etal~\cite{GGP} and may be seen as a protocol between two polynomial-time parties, a \emph{client}, $C$, and a \emph{server}, $S$.
A successful run of the protocol results in the provably correct computation of $F(x)$ by the server for an input $x$ supplied by the client. More specifically, a $\VC$ scheme comprises the following steps~\cite{GGP}:
\begin{enumerate}
  \item  \ifnum\lncs>0 {\sf KeyGen} (\emph{Run once}): \fi $C$ computes evaluation information $EK_{F}$ that is given to $S$ to enable it to compute $F$ \ifnum\lncs=0 (pre-processing) \fi
  \item \ifnum\lncs>0 {\sf ProbGen} (\emph{Run multiple times}): \fi $C$ sends the encoded input $\sigma_{x}$ to $S$ \ifnum\lncs=0 (input preparation) \fi
  \item \ifnum\lncs>0  {\sf Compute} (\emph{Run multiple times}): \fi  $S$ computes $y=F(x)$ using $EK_{F}$ and $\sigma_{x}$ and returns an encoding of the output $\sigma_{y}$ to $C$ \ifnum\lncs=0 (output computation) \fi
  \item \ifnum\lncs>0 {\sf Verify}  (\emph{Run multiple times}): \fi $C$ checks whether $\sigma_{y}$ encodes $F(x)$ \ifnum\lncs=0 (verification) \fi
\end{enumerate}
\ifnum\lncs=0
The operation of a \VC\ scheme is illustrated in Figure~\ref{fig:basic-vc}.
Step 1 is performed once; steps 2--4 may be performed many times.
Step 1 may be computationally expensive but the remaining operations should be efficient for the client.
In other words the cost of the setup phase (to the client) is amortized over multiple computations of $F$.
A \VC\ scheme comprises four algorithms -- {\sf KeyGen}, {\sf ProbGen}, {\sf Compute} and {\sf Verify} -- corresponding to the four steps described above.

\begin{figure}[t]
  \begin{subfigure}[b]{.35\textwidth}\centering
  \begin{tikzpicture}[scale=0.85,transform shape]%
    \tikzstyle{server}=[circle, draw,minimum width=24pt]
    \tikzstyle{entity}=[circle,draw,minimum width=24pt]	
    \node[server](s1){$S$};
    \node[entity, left=2cm of s1](c1){$C$};
    \draw[->] (c1.north east) to[out=25,in=155] node[auto] {\footnotesize 1. $EK_{F}$} (s1.north west);
    \draw[->] (c1) to node[fill=white] {\footnotesize 2. $\sigma_{x}$} (s1);
    \draw[->] (s1.south west) to[out=205,in=335] node[auto] {\footnotesize 3. $\sigma_{y}$} (c1.south east);
    \draw[->] (c1) edge [out=100,in=170,loop] node[auto] {\footnotesize 4.} ();
  \end{tikzpicture}
  \caption{A VC system}\label{fig:basic-vc}
  \end{subfigure}
\hfill
 \begin{subfigure}[b]{.6\textwidth}\centering
  \begin{tikzpicture}[scale=0.85,transform shape]%
    \tikzstyle{server}=[circle, draw,minimum width=24pt]
    \tikzstyle{entity}=[circle,draw,minimum width=24pt]	
    \node[server](s1){$S$};
    \node[entity, left=2cm of s1](c1){$C_1$};
    \node[entity, right=2cm of s1](c2){$C_2$};
    \node[cloud, draw,cloud puffs=11,cloud ignores aspect,below=1.5cm of s1] (public) {Public};
    \draw[->] (c1.north east) to[out=25,in=155] node[auto] {\footnotesize $EK_{F}$} (s1.north west);
    \draw[->] (c1) to node[fill=white] {\footnotesize $\sigma_{x_1}$} (s1);
    \draw[->] (s1.south west) to[out=205,in=335] node[auto] {\footnotesize $\sigma_{y_1}$} (c1.south east);
    \draw[->] (c2) to node[fill=white] {\footnotesize $\sigma_{x_2}$} (s1);
    \draw[->] (s1.south east) to[out=335,in=205] node[auto,swap] {\footnotesize $\sigma_{y_2}$} (c2.south west);
    \draw[->] (c1.south) to[out=270,in=135] node[auto,swap] {\footnotesize $PK_{{F}},VK_{F,x_1}$} (public.puff 2);
    \draw[->] (c2.south) to[out=270,in=45] node[auto] {\footnotesize $VK_{F,x_2}$} (public.puff 11);
    \draw[->] (public) edge [out=30,in=330,loop] node[auto] {\footnotesize    Verify} ();
  \end{tikzpicture}%
    \caption{A PVC system}\label{fig:basic-pvc}
  \end{subfigure}
  \caption{The operation of verifiable computation schemes}
\end{figure}
\fi

Parno \etal{}~\cite{PRV} introduced the idea of \emph{Publicly Verifiable Computation} (PVC).
\ifnum\lncs=0
The operation of a \PVC\ scheme is illustrated in Figure~\ref{fig:basic-pvc}
\fi
In this setting, a single client $C_1$ computes $EK_{F}$, as well as publishing 
information $PK_{{F}}$ that enables other clients to encode inputs, meaning that only one client has to run the expensive pre-processing stage.
Each time a client submits an input $x$ to the server, the client may publish $VK_{F,x}$, which enables any other client to verify that the output is correct.
A PVC scheme uses the same four algorithms as VC but {\keygen} and {\probgen} are now required to output public values that  other clients may use to encode inputs and verify outputs, respectively.

\ifnum\lncs=0
	\input{BackgroundABE}
\fi
\ifnum\lncs=0
\subsection{PVC using Key-Policy Attribute-based Encryption}
\else
\subsubsection{PVC using KP-ABE.}
\fi
\label{sect:parnobackground}
Parno \etal{} provide a concrete instantiation
\ifnum\lncs=0
 of PVC for the case when $F$ is a Boolean function~\cite{PRV}.
 \else
using \emph{Key-policy Attribute-based Encryption}\footnote{If input privacy is required then a predicate encryption scheme could
be used in place of the KP-ABE scheme.} (KP-ABE)~\cite{Waters}, for Boolean functions~\cite{PRV}.
\fi
 \ifnum\lncs=0
Their instantiation makes use of 
	\emph{Key-policy Attribute-based Encryption}\footnote{If input privacy is required then a predicate encryption scheme could
be used in place of the KP-ABE scheme.} (KP-ABE)~\cite{Waters}.
In KP-ABE, each private key is associated with some family of attribute sets $\accessstruct = \{A_1,\dots,A_m\}$, while each ciphertext is computed using a single public key and associated with a single subset of attributes $A$.
Decryption succeeds if the private key includes the attribute set under which the message was encrypted: that is $A_i = A$ for some $i$.
The set of attribute sets defining a private key is usually called an \emph{access structure} and, in most schemes, is \emph{monotonic}, meaning $A' \in \accessstruct$ whenever there exists $A \subset A'$ such that $A \in \accessstruct$.

To instantiate PVC using KP-ABE, we define
\fi
\ifnum\lncs>0
Define
\fi
 a universe $\U$ of $n$ attributes and associate $V \subseteq \U$ with a binary $n$-tuple in which the $i$th place is $1$ if and only if the $i$th attribute is in $V$.
We call this the \emph{characteristic tuple} of $V$.
Thus, there is a natural one-to-one correspondence between $n$-tuples and attribute sets; we write $A_x$ to denote the set associated with $x$.
A function $F : \{0,1\}^n \rightarrow \{0,1\}$ is monotonic if $x \leqslant y$ implies $F(x) \leqslant F(y)$, where $x = (x_1,\dots,x_n)$ is less than or equal to $y = (y_1,\dots,y_n)$ if and only if $x_i \leqslant y_i$ for all $i$.
For a monotonic function 
\ifnum\lncs=0
$F : \{0,1\}^n \rightarrow \{0,1\}$, 
\else
F, 
\fi
the set $\{x \in \{0,1\}^n : F(x) = 1\}$ defines a monotonic access structure which we denote  $\accessstruct_F$.

\begin{table}[t]\centering
\caption{PVC using KP-ABE}\label{tbl:pvc-using-kp-abe}
  \begin{tabular}{ll}
  \toprule
    \bf Abstract PVC parameter~~ & \bf Parameter in KP-ABE instantiation \\
  \midrule
  $EK_{F}$ & $SK_{\accessstruct_F}$ \\ %
    $PK_{F}$ & Master public key $\PP$ \\
    $\sigma_{x}$ & Encryption of $m$ using $\PP$ and $A_x$ \\
    $\sigma_{y}$ & $m$ or $\bot$ \\
    $VK_{F,x}$ & $g(m)$ \\
  \bottomrule
  \end{tabular}
\end{table}

The mapping between PVC and KP-ABE parameters are shown in Table~\ref{tbl:pvc-using-kp-abe}.
Informally, for a Boolean function $F$, the client generates a
private key $SK_{\accessstruct_F}$ using the KP-ABE {\sf KeyGen} algorithm.
Given an input $x$, a client encrypts a random message $m$ ``with''
$A_x$ using the KP-ABE {\sf Encrypt} algorithm and publishes $VK_{F,x} = g(m)$
where $g$ is a suitable one-way function (\eg\ a hash function).
The server decrypts the message using the KP-ABE {\sf Decrypt} algorithm, which will either return $m$ (when $F(x) = 1$) or $\bot$.
The server returns $m$ to the client.
Any client can test whether the value returned by the server is equal to $g(m)$.
Note, however, that a ``rational'' malicious server will always return $\bot$, since returning any other value will (with very high probability) result in the verification algorithm returning a reject decision.
Thus, it is necessary to have the server compute both $F$ and its ``complement'' (and for both outputs to be verified).
We revisit this point in 
\ifnum\lncs=0
Sect.~\ref{sect:construction}. 
\else
Appendix~\ref{sect:construction}.
\fi
The interested reader may also consult the original paper for further details~\cite{PRV}.
\ifnum\lncs=0

\fi
Note that in order to compute the private key $SK_{\accessstruct_F}$, it is necessary to identify all minimal elements $x$ of $\{0,1\}^n$ such that $F(x) = 1$.
There may be exponentially many such $x$.
Thus, the initial phase is indeed computationally expensive for the client.
Note also that the client may generate different private keys to enable the evaluation of different functions.

\ifnum\lncs>0
\subsubsection{Other Related Work.}

\label{sect:related}
The concept of \emph{non-interactive} verifiable computation was formalised by Gennaro \etal{}~\cite{GGP} who gave a construction using Garbled Circuits~\cite{DBLP:conf/focs/Yao86} with fully homomorphic encryption \cite{DBLP:conf/stoc/Gentry09} to re-randomise the circuit to allow multiple executions. 
In independent and concurrent work, Carter \etal~\cite{cart:white14} introduce a third party to generate garbled circuits for such schemes but require this entity to be online throughout and model the system as a secure multi-party computation between the client, server and third-party.
Some works~\cite{gold:multi13,CKKC} consider the multi-client case where functions are computed over joint input from multiple clients and notions such as input privacy become more important. 
\else
\subsection{Other Related Work}

\label{sect:related}
Gennaro \etal{}~\cite{GGP} formalized the problem
of \emph{non-interactive} verifiable computation in which there is only one round of
interaction between the client and the server each time a computation is
performed and introduced a construction based on Yao's Garbled Circuits
\cite{DBLP:conf/focs/Yao86} which provides a ``one-time'' \VOC\ allowing a client to outsource the evaluation of a function on a single input. However it is insecure if the circuit is reused on a different input and thus this cost cannot be amortized, and the cost of generating a new garbled circuit is approximately equal to the cost of evaluating the function itself. To overcome this, the authors additionally use a fully homomorphic encryption scheme \cite{DBLP:conf/stoc/Gentry09} to re-randomize the garbled circuit for multiple executions on different inputs. In independent and concurrent work, Carter \etal~\cite{cart:white14} introduce a third party to generate garbled circuits for such schemes but require this entity to be online throughout the computations and models the system as a secure multi-party computation between the client, server and third-party.
\ifnum\lncs=0
We do not believe this solution is practical in all situations since it is conceivable that a trusted entity is not always available to take part in computations, for example in the battlefield scenario discussed in Section~\ref{sect:intro}. Here, the \caa\ could be physically located within a high security base or governmental building and field agents may receive relevant keys before being deployed, but actual computations are performed using more local available servers and communications links. It may not be feasible, or desirable, for a remote agent to contact the headquarters and maintain a communications link with them for the duration of the computation. In addition, the \caa\ could easily become a bottleneck in the system and limit the number of computations that can take place at any one time, since we assume there are many servers but only a single (or small number of) trusted third parties.
\fi

Some works have also considered the multi-client case in which the input data to be sent to the server is shared between multiple clients, and notions such as input privacy become more important. 
Choi \etal~\cite{CKKC} extended the garbled circuit approach~\cite{GGP} using a proxy-oblivious transfer primitive to achieve input privacy in a
non-interactive scheme. Recent work of Goldwasser \etal~\cite{gold:multi13} extended the construction of Parno \etal~\cite{PRV} to allow multiple clients to provide input to a functional encryption algorithm.
\fi

%% file: SCdefs.tex
\section{PVC with a Key Distribution Center}
\label{sect:ca}
\label{sect:scdefs}

We now introduce our extension of PVC, which we call \emph{Publicly Verifiable Computation with a Key Distribution Center} (\PVCCA).
We assume there are many clients and multiple servers.
Different servers may compute the same function $F$ and servers are ``certified'' to compute $F$ by the \ca.
\ifnum\lncs=0

\fi
As we briefly explained in the introduction, there appear to be good reasons for adopting an architecture of this nature and several scenarios in which such an architecture would be appropriate.
\ifnum\lncs=0
The increasing popularity of relatively lightweight mobile computing devices in the workplace means that complex computations may best be performed by more powerful servers run by the organization.
One can also imagine clients delegating computation to servers in the cloud and would wish to have some guarantee that those servers are certified to perform certain functions.

\else
The increasing popularity of relatively lightweight mobile computing devices in the workplace means that complex computations may best be performed by more powerful servers run by the organization or in the cloud and we would wish to have some guarantee that those servers are certified to perform certain functions.
\fi
It is essential that we can verify the results of the computation.
If cloud services are competing on price to provide ``computation-as-a-service'' then it is important that a server cannot obtain an unfair advantage by simply not bothering to compute $F(x)$ and returning garbage instead.
It is also important that a server who is not certified cannot return a result without being detected.
\ifnum\lncs=0
We examine the possible security concerns arising from \PVCCA\ in Sect.~\ref{sect:security}.

\fi
\begin{figure}[t]
  \centering
   {\fontsize{8}{11}\selectfont
 \begin{tabular}{l p{.1\textwidth} p{.15\textwidth} p{.22\textwidth}}
 \toprule
  \multirow{2}{*}{\bf Algorithm~~} & \multicolumn{3}{c}{\bf Run by} \\
 \cmidrule{2-4}
   & \bf VC~~ & \bf PVC~~ & \bf \PVCCA \\
 \midrule
  {\sf KeyGen} & $C_1$ & $C_1$ & \caa \\
  {\sf ProbGen} & $C_1$ & $C_1,C_2, \dots$ &  $C_1,C_2,\dots$ \\
  {\sf Compute} & $S$ &  $S$ &  $S_1,S_2, \dots$ \\
  {\sf Verify} & $C_1$ & $C_1,C_2,\dots$ &  $C_1,C_2,\dots$ or $M$ \\
 \bottomrule
 \end{tabular}
 }
\hfill
   \begin{subfigure}[t]{0.49\textwidth} \centering
\vspace{0pt}
\hspace{-52.5pt}
  \scalebox{.75}{
 \begin{tikzpicture}
    \tikzstyle{server}=[circle, draw,minimum width=24pt]
    \tikzstyle{entity}=[circle,draw,minimum width=24pt]	
    \node[server](ttp){$\caa$};
    \node[server,left=1.4cm of ttp](s1){$S_1$};
    \node[server,right=1.45cm of ttp](s2){$S_2$};
    \node[server,right=0.7cm of s2](s3){$S_3$};
    \node[cloud, draw,cloud puffs=11,cloud ignores aspect,below=1.3cm of ttp] (public) {Public};
    \node[entity, left=1.1cm of public](c1){$C_1$};
    \node[entity, right=1.15cm of public](c2){$C_2$};
    \draw[->] (ttp) to node[auto,swap] {\footnotesize $EK_{F, S_1}$} (s1);
    \draw[->] (ttp) to node[auto] {\footnotesize $EK_{F, S_2}$} (s2);
    \draw[->] (ttp.north east) to[out=30,in=150] node[auto] {\footnotesize $EK_{G, S_3}$} (s3.north west);
    \draw[->] (c1) to node[auto] {\footnotesize $\sigma_{x_1}$} (s1);
    \draw[->] (s1) to node[auto] {\footnotesize $\sigma_{y_1}$} (c1);
    \draw[->] (c2) to node[auto,pos=0.7] {\footnotesize $\sigma_{x_2}$} (s2);
    \draw[->] (s2) to node[auto,pos=0.3] {\footnotesize $\sigma_{y_2}$} (c2);
    \draw[->] (c2.east) to[out=0,in=270] node[auto] {\footnotesize $\sigma_{x_3}$} (s3.south);
    \draw[->] (s3.south) to[out=270,in=0] node[auto] {\footnotesize $\sigma_{y_3}$} (c2.east);
    \draw[->] (c1) to node[auto,swap] {\footnotesize $VK_{F,x_1}$} (public);
    \draw[->] (public) to node[auto,swap] {} (c1);
    \draw[->] (c2) to node[auto, swap] {\footnotesize $VK_{F,x_2}$} (public);
    \draw[->] (public) to node[auto,swap] {} (c2);
    \draw[->] (c2.south west) to[out=210,in=-30] node[auto, pos=0.55] {\footnotesize $VK_{G,x_3}$} (public.south east);
        \draw[<-] (c2.south west) to[out=210,in=-30, swap] node[auto] {} (public.south east);

        \draw[->] (public) to node[auto, pos=0.3] {\footnotesize Revoke} (ttp);
    \draw[->] (ttp) to node[auto,pos=0.7] {\footnotesize $PK_{F}, PK_G$} (public);
    \draw[->] (c1) edge [out=320,in=220,loop, swap] node[auto] {\footnotesize
        Verify} (c1);
    \draw[->] (c2) edge [out=250,in=320,loop,swap] node[auto, pos=0.8] {\footnotesize
                Verify} (c2);
  \end{tikzpicture}
  }
  \caption{Standard Model}
  \end{subfigure}
  \hfill
\begin{subfigure}[t]{0.49\textwidth} \centering
\vspace{5pt}
  \scalebox{.75}{
 \begin{tikzpicture}
    \tikzstyle{server}=[circle, draw,minimum width=24pt]
    \tikzstyle{entity}=[circle,draw,minimum width=24pt]	
           \node[server](ttp){$\caa$};

   \node[server, left=of ttp](s1){$S_1$};
           \node[server,below=of ttp](man){$M$};

      \node[server,below=of s1](s2){$S_2$};
    \node[cloud, draw,cloud puffs=11,cloud ignores aspect,right=1.3cm of ttp] (public) {Public};

      \node[entity, below=of public](c1){$C_1$};
            \node[entity, below=of c1](c2){$C_2$};

    \draw[->] (ttp) to node[auto,swap, near start] {\footnotesize $EK$} (s1);
    \draw[->] (ttp) to node[auto,swap, near end] {\footnotesize $EK$} (s2);
    \draw[->] (c1) to node[auto] {\footnotesize $\sigma_{x_1}$} (man);
        \draw[->] (man) to node[auto] {} (s1);
   \draw[->] (s1) to node[auto, pos=0.2] {\footnotesize $\sigma_{y_1}$} (man);
      \draw[->] (man) to node[auto] {\footnotesize $\tau_{\sigma_{x_1}}$} (c1);
 \draw[->] (c2) to node[auto, pos=0.3, sloped] {\footnotesize $\sigma_{x_2}$} (man);
        \draw[->] (man) to node[auto] {} (s2);
   \draw[->] (s2) to node[auto, near start] {\footnotesize $\sigma_{y_2}$} (man);
      \draw[->] (man) to node[auto, pos=0.5, sloped] {\footnotesize $\tau_{\sigma_{x_2}}$} (c2);
       \draw[->] (c1) to node[auto] {\footnotesize $VK$} (public);
\draw[->] (public) to node[auto,swap] {} (c1);
\draw[->] (ttp) to node[auto,pos=0.5] {\footnotesize $PK$} (public);
\draw[->] (c2) edge [out=15,in=330] node[auto, swap] {\footnotesize $VK$} (public);
    \draw[->] (public) edge [out=330,in=15]node[auto,swap] {} (c2);
     \draw[->] (man) to node[auto, pos=0.6,swap] {\footnotesize Revoke} (ttp);
     \draw[->] (man) edge [out=220,in=320,loop] node[auto] {\footnotesize BVerify} (man);
     
     \draw[->] (c1) edge [out=320,in=220,loop, swap] node[auto] {\footnotesize
             ROut} (c1);
     \draw[->] (c2) edge [out=320,in=220,loop, swap] node[auto] {\footnotesize
                  ROut} (c2);        

     \node[fit=(s1) (s2)] (s12) {};
\node[fit=(s2) (man)] (s2m) {};
\draw[dashed] (s12.north west) -- ($(s12.north east) - (0.25,0)$-| s2m.north) -- ($(s2m.north east) - (0, 0.5)$) -- (s2m.south east) -- (s2m.south west) -- cycle;

  \end{tikzpicture}
  }
  
  \vspace{-9pt}
 \caption{Manager model}
  \end{subfigure}
  \hfill

    \caption{The operation of \PVCCA}
  \label{fig:pvc-ttp}
\end{figure}
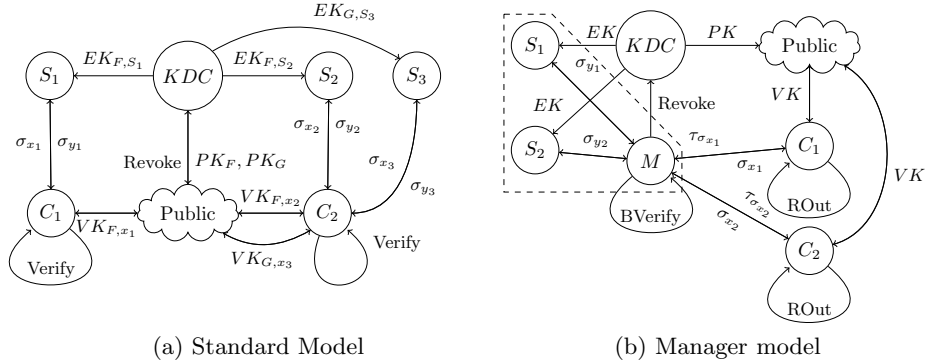
\ifnum\lncs=0
The basic idea of our scheme is to have the \caa\ perform the expensive setup operation.
The \caa\ provides each server with a distinct key to compute $F$.
A client may request the computation of $F(x)$ from any server that is certified to compute $F$, as illustrated in Fig.~\ref{fig:pvc-ttp}.

\fi
In this paper we focus on two example system architectures, which we call the Standard Model and the Manager Model:
\begin{itemize}[noitemsep,topsep=-0pt]
\item The standard model is a natural extension of the \PVCa\ architecture with the addition of a \caa.
The entities comprise a set of clients, a set of servers and a \caa.
The \caa\ initialises the system and generates keys to enable verifiable computation.
Clients submit computation requests to a particular server and publish some verification information.
Any party can verify the correctness of a server's output.
If the output is incorrect, the client may report the server to the the \caa\ for revocation which will prevent the server from performing any further computations of this function.
\item The manager model, in contrast, employs an additional Manager entity who ``owns'' a pool of computation servers.
Clients submit jobs to the manager, who will select a server from the pool based on workload scheduling, available resources or as a result of some bidding process if servers are to be rewarded per computation.
A plausible scenario is that  servers enlist with a manager to ``sell'' the use of spare resources, whilst clients subscribe to utilise these through the manager.
Results are returned to the manager who should be able to verify the server's work.
The manager forwards correct results to the client whilst a misbehaving server may be reported to the \caa\ for revocation, and the job assigned to another server.
Due to public verifiability, any party with access to the output and the verification token can also verify the result.
However, in many situations we may not desire external entities to access the result, yet there remains legitimate reasons for the manager to perform verification.
Thus we introduce ``blind verification'' such that the manager (or other entity) may verify the validity of the computation without learning the output, but the delegating client holds an extra piece of information that enables the output to be retrieved.
\end{itemize}
A \PVCCA\ system operates as follows\ifnum\lncs>0, and as shown in Figure~\ref{fig:pvc-ttp}:
\begin{enumerate}
 \item The \caa\  initializes the system and generates public and private parameters, which are used
       to generate new keys for many different functions.
       \item A server $S$ may join the system by registering with the \caa\ to receive a private key $SK_S$. 
 \item A server $S$ that wishes to provide a computation service for a
      function $F$  makes a request to the \caa, who generates a
     (personalised) \emph{secret} value $EK_{{F, S}}$ and transmits it
     to $S$; this will be used in the computation of $F(x)$.\footnote{Note
     that for the purposes of revocation $EK_{{F, S}}$ is associated to
     a particular server $S$ and is not public as in previous schemes.}
 \item The \caa\ also generates public data $PK_{{F}}$ for  $F$, used to encode client inputs, and publishes a list of servers, $L_{{F}}$, that are certified to compute $F$.
 \item To outsource the computation of $F(x)$, a client $C$ uses $PK_{{F}}$ to prepare $\sigma_{x}$ and public verification value $VK_{{F,x}}$. In the standard model, the client sends $\sigma_{x}$ to a selected server; in the manager model, $\sigma_{x}$ is sent to a manager who distributes it to a server according to some policy or bidding process. 
  \item On receipt of a request to evaluate $F(x)$, (an honest server) $S$ computes the encoded output $\sigma_{y}$ using $EK_{{F, S}}$ and $\sigma_{x}$, publishing or returning $\sigma_{y}$ to $C$ in the standard model, or to the manager in the manager model.
 \item In the standard model, any party, $V$, can run a verification algorithm using $\sigma_{y}$ and $VK_{{F,x}}$ as inputs; the algorithm will output \accept\ and $F(x)$ if and only if  $\sigma_y$ encodes $F(x)$.  
 \ifnum\lncs=0
 Thus $V$ learns the result and whether $S$ correctly evaluated the function or otherwise misbehaved.
 \fi
 In the manager model, the manager (or other) runs a blind verification algorithm such that they learn whether the output is valid, but \emph{not} the actual value of $F(x)$ -- thus the manager learns whether to reward or request the revocation of the server and may enlist an additional server if required, but does not learn the (sensitive) result. The delegator or other chosen verifiers hold additional information that can be  used in an output retrieval algorithm to learn the value of $F(x)$ without having to verify the result and potentially resubmit. These steps could be run together as in the standard model, and we collectively call these the verification algorithm.
 \item If an invalid result is detected, the verifier or manager reports $S$ by sending a token $\tau_{\sigma_y}$ to the \caa, who will revoke $S$. Thus $S$ may incur a financial penalty from being unable to compute $F$  until the \caa\ re-certifies him.
\end{enumerate}

\subsection{Formal Details}
\ifnum\lncs=0
We now present a more formal definition of the algorithms involved in a \PVCCA\
scheme.
\begin{definition}%
    \label{def:pvcca}
{\rm
    A  \emph{Publicly Verifiable Outsourced Computation Scheme with
    Key Distribution Center (\PVCCA)} comprises the following algorithms:

    \begin{itemize}

        \item{$\setup(1^\lambda) \rightarrow (\PP, \MSK)$:}
                  The \setup\ algorithm is run by the \caa\ to establish the public
            parameters \PP\ and a master secret key $\MSK$.

        \item{$\fninit(\PP, \MSK, F) \rightarrow (PK_F, L_{{F}})$:}
                    The \fninit\ algorithm is run by the \caa\ to generate a public
            delegation key, $PK_F$, for a function $F$ as well as a list
            $L_{{F}}$ of available servers for evaluating $F$, which is
            initially empty.
            
           \item{$\register(\PP, \MSK, S) \rightarrow SK_S$:}
                     The \register\ algorithm is run by the \caa\ to generate a personalised key $SK_S$ for a computation server $S$.

        \item{$\certify(\PP, \MSK, F, L_{{F}}, S) \rightarrow
            (EK_{{F,S}}, L_{{F}})$:}
                    The \certify\ algorithm is run by the \caa\ to generate a certificate
            in the form of an evaluation key $EK_{{F,S}}$ for a function $F$ and
            server $S$. The server $S$ is added to the list, $L_{{F}}$, of available
            servers for evaluating $F$.

        \item{$\probgen(x, PK_F) \rightarrow (\sigma_x, VK_{F,x})$:}
                    The \probgen\ algorithm is run by a client to delegate the
            computation of $F(x)$ to a server.

        \item{$\compute(\sigma_x, EK_{{F,S}}, SK_S) \rightarrow \sigma_y$:}
                    The \compute\ algorithm is run by a server $S$ in possession of an
            evaluation key $EK_{{F,S}}$, $SK_S$ and an encoded input $\sigma_x$ of $x$ to evaluate $F(x)$ and output an encoding,
            $\sigma_y$, of the  result, which includes an identifier of $S$.

         \item{$\verify(\PP, \sigma_y, VK_{F,x}, L_{{F}}) \rightarrow (\tilde{y},
                        \tau_{\sigma_y})$:}
                        The \verify\ algorithm is run by any party in possession of a
                        verification key $VK_{F,x}$ for the evaluation of $F(x)$ to check that
                        the encoded output, $\sigma_y$, returned by the
                        server is a valid encoding of $y = F(x)$.
                        It returns a result $\tilde{y}$ which is either $y$ or $\bot$ (if
                        the server did not perform the computation correctly). 
                        If $\tilde{y} = y$ then the token $\tau_{\sigma_y} = (\accept,S)$ is
                        generated, else   $\tilde{y} = \bot$ and the generated token is
                        $\tau_{\sigma_y} = (\reject, S)$ signifying a misbehaving server.

        \item{$\revoke(\MSK, \tau_{\sigma_y}, F, L_{{F}}) \rightarrow
            (\{EK_{F,S^\prime}\}, L_{{F}}) \textrm{ or } \perp$:}
                        The \revoke\ algorithm is run by the \caa\ if a client reports that a
            server misbehaved \ie\ that \verify\ returned $\tau_{\sigma_y} =
            (\reject, S)$ (if
            $\tau_{\sigma_y} = (\accept,S)$ then this algorithm should output
            $\perp$).
            It revokes the evaluation key $EK_{{F,S}}$ of the server $S$ thereby preventing any further evaluations of $F$. This is achieved by removing $S$ from $L_F$ (the list of servers for $F$) and issuing updated evaluation keys $EK_{F,S^\prime}$ to all servers $S^\prime \neq S$.
            
    \end{itemize}
    }
\end{definition}

\else  %
\begin{definition}
    \label{def:pvcca}
{\rm
    A  \emph{Publicly Verifiable Outsourced Computation Scheme with
    Key Distribution Center (\PVCCA)} comprises the following algorithms:

    \begin{itemize}

        \item{$\setup(1^\lambda) \rightarrow (\PP, \MSK)$:}
                  Run by the \caa\ to establish public
            parameters \PP\ and a master secret key $\MSK$.

        \item{$\fninit(\PP, \MSK, F) \rightarrow (PK_F, L_{{F}})$:}
                    Run by the \caa\ to generate a public
            delegation key, $PK_F$, for a function $F$ as well as a list
            $L_{{F}}$ of available servers for evaluating $F$, which is
            initially empty.
            
           \item{$\register(\PP, \MSK, S) \rightarrow SK_S$:}
                     Run by the \caa\ to generate a personalised key $SK_S$ for a computation server $S$.

        \item{$\certify(\PP, \MSK, F, L_{{F}}, S) \rightarrow
            (EK_{{F,S}}, L_{{F}})$:}
                   Run by the \caa\ to generate a certificate
            in the form of an evaluation key $EK_{{F,S}}$ for a function $F$ and
            server $S$.  $S$ is added to the list, $L_{{F}}$, of available
            servers for evaluating $F$.

        \item{$\probgen(x, PK_F) \rightarrow (\sigma_x, VK_{F,x}, b)$:}
                    The \probgen\ algorithm is run by a client to delegate the
            computation of $F(x)$ to a server. The output value $b$ is used to enable output retrieval after the blind verification step.

        \item{$\compute(\sigma_x, EK_{{F,S}}, SK_S) \rightarrow \sigma_y$:}
                    Run by a server $S$ in possession of an
            evaluation key $EK_{{F,S}}$, $SK_S$ and an encoded input $\sigma_x$ of $x$ to evaluate $F(x)$ and output an encoding,
            $\sigma_y$, of the  result, which includes an identifier of $S$.
            
        \item{$\verify(\PP, \sigma_y, VK_{F,x}, L_{{F}}) \rightarrow (\tilde{y},
            \tau_{\sigma_y})$:}
            Verification consists of two steps.
            
            \begin{itemize}
            \item{$\blindverify(\PP,\sigma_y, VK_{F,x}, L_{{F}}) \rightarrow (\mu, \tau_{\sigma_y})$:}
            	Run by any verifying party party (standard model), or run by the manager (manager model), in  
            	possession of $VK_{F,x}$ and encoded output, $\sigma_y$. This outputs a token $\tau_{\sigma_y} = (\accept, S)$ if the output is valid, or $\tau_{\sigma_y} = (\reject, S)$ if $S$ misbehaved. It also outputs $\mu$ which is an encoding of the actual output value.
	           	
            \item{$\retrieve(\mu, \tau_{\sigma_y}, VK_{F,x}, b) \rightarrow \tilde{y}$:}
            	Run by a verifier in possession of $b$ to retrieve the actual result $\tilde{y}$ which is either $F(x)$ or $\bot$.
            \end{itemize}

        \item{$\revoke(\MSK, \tau_{\sigma_y}, F, L_{{F}}) \rightarrow
            (\{EK_{F,S^\prime}\}, L_{{F}}) \textrm{ or } \perp$:}
                        Run by the \caa\ if a misbehaving server is reported \ie\ that \verify\ returned $\tau_{\sigma_y} =
            (\reject, S)$ (if
            $\tau_{\sigma_y} = (\accept,S)$ then this algorithm should output
            $\perp$).
            It revokes the evaluation key $EK_{{F,S}}$ of the server $S$ thereby preventing any further evaluations of $F$. This is achieved by removing $S$ from $L_F$ (the list of servers for $F$) and issuing updated evaluation keys $EK_{F,S^\prime}$ to all servers $S^\prime \neq S$.
            
    \end{itemize}
    }
\end{definition}
\fi

\ifnum\lncs=0
We say that a \PVCCA\ scheme is \emph{correct} if the verification algorithm
almost certainly outputs \accept\ when run on a valid verification key and an encoded output honestly
produced by a computation server given a validly generated encoded input and
evaluation key. That is, if all algorithms are run honestly then the verifying
party should almost certainly accept the returned result. 
\fi
\ifnum\lncs=0
A more formal definition follows:
\vspace{2cm}
\fi
\begin{definition}[Correctness]
\label{def:pvccasec}
A  Publicly Verifiable Computation Scheme with a Key Distribution Center (\PVCCA) is \emph{correct} for a family of functions $\mathcal{F}$ if for all functions $F \in \mathcal{F}$ and inputs $x$,where \textup{negl}$(\cdot)$ is a negligible function of its input:
\begin{align*}
\Pr[
&\setup(1^\lambda) \rightarrow (\PP, \MSK), \fninit(\PP, \MSK, F) \rightarrow (PK_F, L_{\textrm{F}}),\\
&\register(\PP, \MSK, S) \rightarrow SK_S, \certify(\PP, \MSK, F, L_{\textrm{F}}, S) \rightarrow
            (EK_{\textrm{F,S}}, L_{\textrm{F}}),\\
&\probgen(x, PK_F) \rightarrow (\sigma_x, VK_{F,x},b),\\
&\verify(\PP,\compute(\sigma_x, EK_{F,S}, SK_S), VK_{F,x}, L_{\textrm{F}}) \rightarrow (F(x), (\accept,S))
]\\
&= 1- \textup{negl}(\lambda).
\end{align*}

\end{definition}

\input{SecurityModels}

%% file: SecurityModels.tex
\subsection{Security Models}
\label{sect:security}

\ifnum\lncs=0
We now introduce several security models capturing different requirements of a
\PVCCA\ scheme. We will formalise these notions of security as a series of
cryptographic games run by a challenger.
\else
We now formalise several notions of security as a series of cryptographic games.
\fi
 The adversary against a particular function $F$
is modelled as a probabilistic polynomial time algorithm $\adv$ run by a challenger\ifnum\lncs>0. \else during the game with input parameters chosen to
represent the knowledge of a real attacker as well the security parameter $\lambda$.\fi 
The adversary algorithm may 
\ifnum\lncs=1
maintain state and be multi-stage 
\fi
\ifnum\lncs=0
(\ie\ called several
times by the challenger, with different input parameters)
\fi and we overload the
notation by calling each of these adversary algorithms $\adv$.
\ifnum\lncs=0
This represents the adversary performing tasks at different points during the
execution of the system, and we assume that the adversary may maintain a state
storing any knowledge it gains during each phase (we do not
provide the state as an input or output of the adversary for ease of notation).
\fi
The notation $\adv^\oraclesign$ is used to denote the adversary $\adv$ being
provided with oracle access to the following functions: $\fninit(\PP,
\MSK,\cdot)$,$\register(\PP, \MSK, \cdot)$, $\certify(\PP,
        \MSK,\cdot,\cdot,\cdot)$ and $\revoke(\MSK, \cdot, \cdot, \cdot)$.
        \ifnum\lncs=0
        This means that the adversary can query (multiple times) the challenger for any of these
        functions with the adversary's choice of values for parameters
        represented with a dot above. This models information the adversary
        could learn from observing a functioning system or by acting like a
        legitimate client (or corrupting one) to request some functionality.
        
        \fi
        In each of the games, we define the \emph{advantage} and \emph{security} of $\adv$ as:
        \begin{definition}
The \emph{advantage} of an adversary $\mathcal{A}$
running in probabilistic polynomial time (PPT), making a polynomial number of
queries $q$ is defined as follows, where ${\bf X} \in \{PubVerif, Revocation, VindictiveS, BVerif, VindictiveM\}$:
\begin{center}
 $Adv_{\adv}^{\bf X}(\VCpub, F, 1^\lambda, q) =
 \textup{Pr}[\textup{\textbf{Exp}}_{\adv}^{\bf X}[\VCpub,F,1^\lambda]
 =1]$.
\end{center}
A \PVCCA\
is \emph{secure against Game {\bf X}} for a function $F$, if for all PPT adversaries $\mathcal{A}$,
Adv$_{\mathcal{A}}^{\bf X}$($\VCpub$, $F$, $1^\lambda$,q) $\leq$ \textup{negl}$(\lambda)$.
\end{definition}

\ifnum\lncs=0
The introduction of the \caa\ and subsequent changes in operation give rise to new security concerns:
\begin{itemize}
 \item Since two (or more) servers may be able to compute the same function, it is important to ensure that servers cannot collude in order to convince a client to accept an incorrect output as correct.
 \item We must ensure that neither an uncertified nor a de-certified server can convince a client to accept an output.
 \item We must ensure that a malicious server $S$ cannot convince a client to believe an honest server has produced an incorrect output.
\end{itemize}
\fi
\subsubsection{Public Verifiability.}

In Game~\ref{game:pv} we 
\ifnum\lncs=0
wish
\else
extend the Public Verifiability game of Parno \etal~\cite{PRV}
\fi
 to formalize that multiple servers should not be able to collude to 
\ifnum\lncs=0
gain
an advantage in convincing
\else
convince
\fi
 \emph{any} verifying party of an incorrect output
(\ie\ that \verify\ returns \accept\ on a
$\sigma_y$ for $y \neq F(x)$).
The game begins (line $1$) with the adversary selecting a (polynomially sized) set of $n$ input values that he
would like to see the problem encoding of (and the corresponding time
period\footnote{The time period here is changed every time a server is
revoked. Alternatively, the time period could be regularly updated but the
\revoke\ algorithm must be run at each interval even if the revocation list has
not changed.}). The challenger runs \setup, \fninit\ and
\register\ to initialise the system and create a public delegation key for a function $F$ given
as a parameter to the game (lines $2$ to $4$). The adversary is given the delegation key, his
private key and the public parameters (\ie\ all values known to a
server in the real setting), and must output a list of servers that should
be certified to compute $F$ (line 6)\footnote{This corresponds to the revocation list in
the model of \cite{attr:attr09} except that we consider a certification list of
servers that should receive the update keys rather than a revocation list of
servers that should not receive these keys. The requirement to output this list
here is due to the selective \INDSHRA\ game that we base the construction upon. Since
this is used in a black-box manner however, a stronger primitive may allow this
game to be improved accordingly.}.

The challenger then runs  \probgen\  for each challenge input
and gives the encoded inputs to the adversary.
The adversary also has oracle access 
\ifnum\lncs=0
to the \fninit, \register, \certify\ and
\revoke\
algorithms 
\fi
to model the corruption of other servers (line 10), and aims to create an encoded
output that is accepted by the challenger yet is not valid for
any challenge input.

\begin{gamefloat}
 {\fontsize{8}{11}\selectfont
\begin{algorithmic}[1]
         \STATE $\{t_{i}^{\star}, x_{i}^{\star}\}_{i \in [n]} \leftarrow \adv(1^\lambda)$;
        \STATE $(\PP, \MSK) \leftarrow \setup(1^\lambda)$;
        \STATE $(PK_F, L_{{F}}) \leftarrow \fninit(\PP, \MSK, F)$;
        \STATE $SK_{\adv} \leftarrow \register(\PP, \MSK, \adv)$;
        \STATE $L_{{F}} \leftarrow \adv(PK_F, \PP, L_{{F}}, SK_{\adv})$; 
                \STATE $(EK_{F, \adv}, L_F) \leftarrow \certify(\PP, \MSK, F, L_{{F}}, \adv)$;
          \FOR{$i=1$ \TO $n$}
                        \STATE $\{\sigma_{x_{i}^{\star}}, VK_{F,x_{i}^{\star}}, b_i\} \leftarrow \probgen(\{t_{i}^{\star}, x_{i}^{\star}\}, PK_{F})$;
          \ENDFOR	
          \STATE $\sigma_{y^{\star}} \leftarrow \adv^{\oraclesign}(PK_F, \PP, L_{{F}}, \{\sigma_{x_{i}^{\star}}, VK_{F,x_{i}^{\star}}\}, EK_{F, \adv}, SK_{\adv})$;												
             \IF{$\exists i \in [n] \mbox{ s.t. } 
             (((\tilde{y}, \tau_{\sigma_{y^{\star}}}) \leftarrow \verify(\PP,\sigma_{y^{\star}}, VK_{F,x_{i}^{\star}}, L_F))$ 
             \AND $((\tilde{y}, \tau_{\sigma_y}) \neq (\bot, (\reject,\adv)))$
             \AND $(\tilde{y} \neq F(x_{i}^\star)))$}
             	\RETURN{$1$}
	    \ELSE
             	\RETURN{$0$}
	    \ENDIF
\end{algorithmic}
}
\caption{\gametitle{\adv}{PubVerif}{\VCpub, F, 1^\lambda}}%
\label{game:pv}
\end{gamefloat}
\ifnum\lncs=0
\begin{definition}
		\label{def:MultServers}
The \emph{advantage} of an adversary $\mathcal{A}$
running in probabilistic polynomial time (PPT), making a polynomial number of
queries $q$ is defined as:
\begin{center}
 $Adv_{\adv}^{PubVerif}(\VCpub, F, 1^\lambda, q) =
 \textup{Pr}[\textup{\textbf{Exp}}_{\adv}^{PubVerif}[\VCpub,F,1^\lambda]
 =1]$.
\end{center}
A \PVCCA\
is \emph{secure in the sense of public verifiability} for a function $F$, if for all PPT adversaries $\mathcal{A}$,
Adv$_{\mathcal{A}}^{PubVerif}$($\VCpub$, $F$, $1^\lambda$,q) $\leq$ \textup{negl}$(\lambda)$.
\end{definition}
\fi

\ifnum\lncs=0
Note that this game is a generalisation of the Public Verifiability game of
Parno \etal~\cite{PRV} since they consider the case where the
adversary is limited to learning only one evaluation key and one encoded input.
The motivation for this updated game is that there is a now a trusted party
issuing keys to multiple servers who may collude, as opposed to the traditional
model in which the system comprises a single client choosing a single server
to whom to outsource a computation. Thus we allow the adversary to collect
multiple inputs from clients (represented by choosing the set of target inputs)
and to learn multiple evaluation keys for different functions and associated
with different servers (since evaluation keys are server-specific in our
setting to enable per-server revocation).
\fi

\subsubsection{Revocation.}

In Game~\ref{game:rev} we require that if a server is detected as misbehaving
\ifnum\lncs=0
(\ie\ a result for $F(x)$ causes the \verify\ algorithm to output $(\bot,
(\reject, S))$)
\else
(\ie\ \verify\ outputs $(\bot,(\reject, S))$)
\fi
 then any subsequent evaluations of $F$ by $S$ should be rejected.
 \ifnum\lncs=0
The motivation here is that even
\else
Even
\fi
 though we have
outsourced the costly computation and pre-processing stages to
the server and \caa\ respectively, there is still a cost involved in delegating and verifying
a computation. If a
server is known not to be trustworthy then we remove any
incentive for it to attempt to provide an outsourcing service
\ifnum\lncs=0
 for this
function 
\fi
(since it knows the result will not be accepted). In addition, we may like to punish 
\ifnum\lncs=0
and further
disincentivize 
\fi
malicious servers by removing their ability to perform work (and
earn rewards) for a period of time.
Finally, from a privacy perspective, we may not wish to supply input data to a server that
is known not to be trustworthy.
  In this game the adversary chooses the target input
values as before (line 1) but now the evaluation key $EK_{F,\adv}$ that it had access to when
selecting $x$ is revoked (line 8) before the computation is run. 
\ifnum\lncs=0
The remainder of the
game proceeds as in the Public Verifiability game but we require that the
adversary is no longer able to provide \emph{any} result that verifies
correctly (even $F(x)$).
\else
We require that the
adversary is no longer able to provide \emph{any} result that verifies
correctly (even $F(x)$).
\fi

\begin{gamefloat}
 {\fontsize{8}{11}\selectfont
\begin{algorithmic}[1]
       \STATE $\{t_{i}^{\star}, x_{i}^{\star}\}_{i \in [n]} \leftarrow \adv(1^\lambda)$;
        \STATE $(\PP, \MSK) \leftarrow \setup(1^\lambda)$;
\STATE $(PK_F, L_{{F}}) \leftarrow \fninit(\PP, \MSK, F)$;
       \STATE $SK_{\adv} \leftarrow \register(\PP, \MSK, \adv)$;
               \STATE $L_{{F}} \leftarrow \adv(PK_F, \PP, L_{{F}}, SK_{\adv})$; 
\STATE $(EK_{F,\adv}, L_{{F}}) \leftarrow \certify(\PP, \MSK, F, L_{{F}}, \adv)$;
\STATE $\tau^\star  = (\reject,\adv)  \leftarrow \adv^{\oraclesign}(PK_F, \PP, L_{{F}}, SK_{\adv})$;
\STATE $(\{EK_{F,S}\}, L_{{F}}) \leftarrow \revoke(\MSK, \tau^\star,F, L_{{F}})$;
\FOR{$i=1$ \TO $n$}
               \STATE $\{\sigma_{x_{i}^{\star}}, VK_{F,x_{i}^{\star}}, b_i\} \leftarrow \probgen(\{t_{i}^{\star}, x_{i}^{\star}\}, PK_{F})$;
          \ENDFOR
\STATE $\sigma_{y^{\star}} \leftarrow \adv^{\oraclesign}(PK_F, \PP, L_{{F}}, \{\sigma_{x_{i}^{\star}}, VK_{F,x_{i}^{\star}}\}, \{EK_{F,S}\},SK_{\adv})$;								
       \IF{$\exists i \in [n] \mbox{ s.t. } 
             (((\tilde{y}, \tau_{\sigma_{y^{\star}}}) \leftarrow \verify(\PP,\sigma_{y^{\star}}, VK_{F,x_{i}^{\star}}, L_F))$ 
             \AND $((\tilde{y}, \tau_{\sigma_y}) \neq (\bot, (\reject,\adv))))$}
             \RETURN{$1$}
        \ELSE
        		\RETURN{$0$}
	\ENDIF

     \end{algorithmic}
     }
  \caption{\gametitle{\adv}{Revocation}{\VCpub, F, 1^\lambda}}\label{game:rev}
\end{gamefloat}
\ifnum\lncs=0
\begin{definition}
    \label{def:revoc}
The \emph{advantage} of a PPT adversary $\mathcal{A}$
making a polynomial number of
queries $q$ is defined as:
\begin{center}
 $Adv_{\adv}^{Revocation}(\VCpub, F, 1^\lambda, q) =
 \textup{Pr}[\textup{\textbf{Exp}}_{\adv}^{Revocation}[\VCpub,F,1^\lambda]
 =1]$.
\end{center}
A \PVCCA\
is \emph{secure against revoked servers} for a function $F$, if for all PPT adversaries $\mathcal{A}$,
Adv$_{\mathcal{A}}^{Revocation}$($\VCpub$, $F$, $1^\lambda$,q) $\leq$ \textup{negl}$(\lambda)$.
\end{definition}
\fi

\subsubsection{Vindictive Servers.}

The motivation for this notion of security is the manager model where
\ifnum\lncs=0
 a pool of computational servers is available
to accept a `job' from a client but they are abstracted such that the
client does not know or care about the individual server identities.
\else
the client does not know the identities of servers selected from the pool.
\fi
Now, since an invalid result can lead to revocation, this leads to a new threat model (particularly in systems where servers
gain rewards per computation performed) in which a malicious server may return
incorrect results but attribute them to an alternate server ID such that the
(honest) server is  
\ifnum\lncs=0
revoked and the pool of available servers for future
computations is reduced in size, leading to a likely increase in reward for the
malicious server.
\else
revoked, thus reducing the size of the server pool and increasing the future reward for the malicious server.
\fi
In Game~\ref{game:vindS} the adversary must (on lines 9 and 10) output an invalid result $\sigma_{y^{\star}}$ and the ID of a server $\tilde{S}$ that it aims to cause to be revoked. It is provided with the standard oracle access on line 9 and on line 10 additionally with oracle access to \compute\ such that he can see outputs returned by honest servers (\ie\ modelling the adversary submitting computation requests to the system himself), subject to the following constraints:
\begin{enumerate}[noitemsep,topsep=-4pt]
\item   No query was made of the form $\oracle{\register}{\PP, \MSK, \tilde{S}}$;
\item    As above but also no query was made of the form $\oracle{\compute}{\sigma_{x_i^\star}, EK_{F, \tilde{S}}, SK_{\tilde{S}}}$;
\end{enumerate}
 The adversary wins if the \caa\ believes that $\tilde{S}$ returned $\tilde{y}$ and revokes $\tilde{S}$.

\begin{gamefloat}
 {\fontsize{8}{11}\selectfont
\begin{algorithmic}[1]
       \STATE $\{t_{i}^{\star}, x_{i}^{\star}\}_{i \in [n]} \leftarrow \adv(1^\lambda)$;
        \STATE $(\PP, \MSK) \leftarrow \setup(1^\lambda)$;
\STATE $(PK_F, L_{{F}}) \leftarrow \fninit(\PP, \MSK, F)$;
       \STATE $SK_{\adv} \leftarrow \register(\PP, \MSK, \adv)$;
                      \STATE $L_{{F}} \leftarrow \adv(PK_F, \PP, L_{{F}}, SK_{\adv})$; 
\STATE $(EK_{F,\adv}, L_{{F}}) \leftarrow \certify(\PP, \MSK, F, L_{{F}}, \adv)$;
\FOR{$i=1$ \TO $n$}
               \STATE $\{\sigma_{x_{i}^{\star}}, VK_{F,x_{i}^{\star}}, b_i\} \leftarrow \probgen(\{t_{i}^{\star}, x_{i}^{\star}\}, PK_{F})$;
          \ENDFOR
                   \STATE $\tilde{S} \leftarrow \adv^{\oraclesign}(PK_F, \PP, L_{{F}}, \{(\sigma_{x_{i}^{\star}}, VK_{F,x_{i}^{\star}})\}, EK_{F, \adv}, SK_{\adv})$  subject to Condition 1;
         \STATE $\sigma_{y^{\star}} \leftarrow \adv^{\oraclesign, \compute}(PK_F, \PP, L_{{F}}, \{(\sigma_{x_{i}^{\star}}, VK_{F,x_{i}^{\star}})\}, EK_{F, \adv}, SK_{\adv})$ subject to Condition 2;
          \IF{$\exists i \in [n]$ \mbox{ s.t. } 
          $((\tilde{y}, \tau_{\sigma_{y^{\star}}}) \leftarrow \verify(\PP,\sigma_{y^{\star}}, VK_{F,x_{i}^{\star}}, L_F))$
             \AND $((\tilde{y}, \tau_{\sigma_y}) = (\bot, (\reject,\tilde{S})))$
             \AND $(\perp \nleftarrow \revoke(\MSK, \tau_{\sigma_y}, F, L_F)))$}
             \RETURN{$1$}
             \ELSE
             \RETURN{$0$}
             \ENDIF
\end{algorithmic}
}
\caption{\gametitle{\adv}{VindictiveS}{\VCpub, F, 1^\lambda}}\label{game:vindS}
\end{gamefloat}
\ifnum\lncs=0
\begin{definition}
    \label{def:vind}
The \emph{advantage} of a PPT adversary $\mathcal{A}$
making a polynomial number of
queries $q$ is defined as:
\begin{center}
 $Adv_{\adv}^{Vindictive}(\VCpub, F, 1^\lambda, q) =
 \textup{Pr}[\textup{\textbf{Exp}}_{\adv}^{Vindictive}[\VCpub,F,1^\lambda]
 =1]$.
\end{center}
A \PVCCA\
is \emph{secure against vindictive servers} for a function $F$, if for all PPT adversaries $\mathcal{A}$,
Adv$_{\mathcal{A}}^{Vindictive}$($\VCpub$, $F$, $1^\lambda$,q) $\leq$ \textup{negl}$(\lambda)$.
\end{definition}
\fi

\subsubsection{Vindictive Manager and Blind Verification.}
\label{sect:vindm}
In Appendix \ref{sect:appgames} we give two more security games for the manager model, namely the notions of Vindictive Managers and Blind Verification. Vindictive Managers captures that a manager, being an intermediary in the verification process, may try to accept an incorrect answer in order to convince the client (or other recipient) of incorrect results -- in a simple case, a vindictive manager could simply return the value of the ``result'' to the client.
In the game, we provide the adversary with an encoded output, and require him to output an incorrect result, $\mu$, with an acceptance token which will be accepted by a verifier in \vcretrieve. 
Thus, the goal is essentially to forge an encoded output $\mu$.

The Blind Verification game captures a (weak) notion of output privacy in that it prevents verifiers from learning the output unless they hold an additional key, $b$. It does not prevent the servers themselves learning the output during \compute\ as usually considered output privacy. The challenger selects a random input $x$ and gives the adversary the encoded output of the computation. The adversary must guess $F(x)$.

%% file: Construction.tex
\section{Construction}
\label{sect:construction}
\label{sect:abe}
\ifnum\lncs=0
\subsection{Introduction}
\fi
We now provide an instantiation of a \PVCCA\ scheme.
Our construction is based on that used by Parno \etal~\cite{PRV} (summarised in Sec.~\ref{sect:parnobackground}) which uses
Key-Policy Attribute-based Encryption (KP-ABE) in a black-box manner to outsource the
computation of a Boolean
\ifnum\lncs>0 %
function\footnote{
\else
function. 
\fi
Following Parno \etal\ we restrict our attention to Boolean functions, and in
particular the complexity class $NC^1$ which includes all circuits of depth
$\bigoh(\log n)$.
This class includes common functions of interest such as AND, OR, NOT, equality and
comparison operators, arithmetic operators and regular expressions.
\ifnum\lncs>0
}.
Notice that to achieve the outsourced evaluation of functions with $n$ bit
outputs, it is possible to evaluate $n$ different functions, each of which
applies a mask to output the single bit in position $i$.
\ifnum\lncs=0

Notice also that different function families will require different
constructions from that presented here for Boolean functions. As a trivial
example, verifiable outsourced evaluation of the identity function may only
require the server to sign the input. On the other hand, despite it seemingly
being a natural choice for outsourcing, it is not clear how a \VC\ scheme for NP-complete
problems could be instantiated. A solution for such problems is by definition
difficult to find so should be outsourced, whilst a candidate solution can be
verified efficiently. However, a malicious server could simply return that a
solution cannot be found for the given problem instance, and the restricted
client could not verify the correctness of this statement.
\fi

Recall that if $\perp$ is returned by the server then the verifier is unable to determine whether $F(x)=0$ \emph{or} whether the server misbehaved.
To avoid this issue, we follow Parno {\em et al.} and restrict the family of functions $\mathcal{F}$ we can evaluate to be the set of Boolean functions closed under complement.
That is, if $F$ belongs to $\mathcal{F}$ then $\overbar{F}$, where $\overbar{F}(x) = F(x) \xor 1$, also belongs to $\mathcal{F}$.
Then, the client encrypts two random messages $m_0$ and $m_1$.
The server is required to return the decryption of those ciphertexts.
Thus, a well-formed response satisfies the following:%
\begin{equation*}
  (d_0, d_1) =\begin{cases}
    (m_0, \perp), & \text{if $F(x) = 1$};\\
    (\perp, m_1), & \text{if $F(x) = 0$}.
  \end{cases}
\end{equation*}
Hence, the client will be able to detect whether the server has misbehaved.
\ifnum\lncs=0
\subsection{Technical Details}
\fi
We assume the existence of a \emph{revocable KP-ABE scheme} for a class of functions $\cal F$ that is closed under complement.
Such a scheme defines the algorithms $\ABEsetup$, $\ABEkeygen$, $\ABEkeyupdate$, $\ABEenc$ and $\ABEdec$.
We also make use of a signature scheme with algorithms $\Sig.\keygen$, $\Sig.\sign$ and $\Sig.\verify$ and a pre-image resistant hash function $g$.

Then we construct a publicly verifiable computation scheme for the same class of functions comprising the algorithms $\vcsetup$,
$\vcfninit$, $\vcregister$, $\vccertify$, $\vcprobgen$, $\vccompute$, $\vcverify$ and $\vcrevoke)$.
\ifnum\lncs=0

Note that in the original scheme by Parno~\etal~\cite{PRV} the required security property of the underlying KP-ABE scheme was a one-key \INDCPA\ notion. This is a more relaxed notion that considered in the vast majority of the ABE literature (where usually the adversary is provided with a \keygen\ oracle and the scheme must prevent collusion between holders of different decryption keys). Parno \etal\ could use this property due to their restricted system model where the client certifies for only a single function per set of public parameters (that is, the client must set up a new ABE environment per function). In our setting with a trusted third party however, we are interested in a more decentralised (and more efficient) environment where the \caa\ can issue keys for multiple functions within a single system. Thus we require the more standard, multi-key notion of security usually considered for ABE schemes.

On a similar note, we again mention that the security games presented in this paper are in the selective ABE model and are written in a format that allows consistency with the \INDSHRSS\ game for revocable KP-ABE~\cite{attr:attr09}. However, since the ABE algorithms are used in a black box manner, we believe that choosing a instantiation with stronger security properties (for example, a fully secure ABE~\cite{lewk:full10} scheme supporting indirect user revocation) should easily allow for a correspondingly more secure \VC\ construction as presented here.

Finally, we remark that whereas in the revocable ABE scheme of Attrapadung \etal~\cite{attr:attr09} the update keys were generated for the set of nodes in $Cover(R)$ where $R$ is the list of revoked users (as discussed in Section~\ref{sect:backgroundrevabe}), in our definitions we have a list $L_F$ of certified servers for a particular function.
Thus $L_F$ is the inverse of $R$ \ie\ $L_F = \U_{\rm{id}}\setminus R$. 
To construct the set of nodes for which key update material should be generated, therefore, we can either compute $R$ by taking the complement as above, or by altering the ABE scheme to use the following algorithm instead. First mark all leaves in $L_F$. Then, working from the leaves upwards in a breadth-first manner, mark all nodes that have both children marked and subsequently unmark the child nodes. 
For ease of notation, in the following algorithms we simply associate the roles of $R$ and $L_F$ and pass $L_F$ directly to the revocable KP-ABE algorithms instead of $R$, with the assumption that one of the above transformations has been performed.

\subsection{Construction}
\fi
\ifnum\lncs=0 %
Informally, the scheme operates in the following way. 
\begin{enumerate}
 \item The $\vcsetup$ algorithm establishes public parameters and a master secret key by calling the $\ABEsetup$ algorithm twice.  This algorithm also initializes a list of registered servers $L_{\it Reg}$ and the time period $t$.\label{step:constrsetup}
 \item The $\vcfninit$ algorithm initializes a list of servers $L_F$ authorized to compute function $F$.
 \item The $\vcregister$ algorithm creates a public-private key pair by calling the signature $\keygen$ algorithm. This algorithm is run by the \caa\ and updates $L_{\it Reg}$ to include $S$.
 \item The $\vccertify$ algorithm creates the key $EK_{F,S}$ that will be used by a server $S$ to compute $F$ and calls the $\ABEkeygen$ and $\ABEkeyupdate$ algorithms twice.  The algorithm also updates $L_F$ to include $S$.\label{step:constrkey}
 \item The $\vcprobgen$ algorithm creates a problem instance $(c_0,c_1)$ and a verification key $VK_{F,x}$. This algorithm is run by a client and calls the $\ABEenc$ algorithm twice.\label{step:constrprob}
 \item The $\vccompute$ algorithm is run by a server $S$ and computes $F(x)$.  Given a problem instance $(c_0,c_1)$ it returns $(m_0,\perp)$ if $F(x) = 1$ or $(\perp,m_1)$ if $F(x) = 0$ together with a digital signature computed over the output.
 \item The $\vcverify$ algorithm is run by any verifying party and either accepts the output $(d_0,d_1)$ or rejects it.  This algorithm verifies the signature on the output and then confirms the output is correct using $VK_{F,x}$.
 \item The $\vcrevoke$ algorithm is run by the TTP and redistributes fresh keys to all non-revoked servers.  This algorithm updates $L_F$ and updates $EK_{F,S}$ using the results of two calls to the $\ABEkeyupdate$ algorithm.
\end{enumerate}
\fi
 \ifnum\lncs=0
 We require two distinct sets of system parameters in Step~\ref{step:constrsetup} for the security proof to work. In Step~\ref{step:constrkey} we have to run the \ABE.\keygen\ algorithm twice -- once for $F$ and once for $\overbar{F}$. However, to prevent a trivial win in the \INDSHRSS\ game, the adversary is not allowed to query for a key with a policy that is satisfied by the challenge input attributes. By definition, either $F(x)$ or $\overbar{F}(x)$ will output $1$ and hence one of these will not be able to be queried to the Challenger. Thus we use the two separate parameters such that the non-satisfied function can be queried to the Challenger and the adversary can use the other set of parameters to generate a key himself.
 
 \fi
More formally, our scheme is defined by Algorithms~\ref{alg:vcsetup}--\ref{alg:vcrevoke}.

{\centering
\begin{minipage}{\textwidth}
\begin{algfloat}[H]
{\fontsize{8}{11}\selectfont
 \begin{algorithmic}[1]
  \STATE Let $\U = \U_{\textrm{attr}} \cup \U_{\textrm{ID}} \cup \U_{\textrm{time}}$
  \STATE $(\mpkabe{0}, \mskabe{0}) \leftarrow \ABE.\setup{}(1^\lambda,\U{})$
  \STATE $(\mpkabe{1}, \mpkabe{1}) \leftarrow \ABE.\setup{}(1^\lambda, \U{})$
  \STATE $L_{\textrm{Reg}} = \epsilon$ (\ie\ an empty list is created)
  \STATE Initialize $\tau$
  \STATE $\PP = (\mpkabe{0}, \mpkabe{1}, L_{\textrm{Reg}}, t)$
  \STATE $\MSK = (\mskabe{0}, \mskabe{1})$
 \end{algorithmic}
 }
 \caption{\vcsetup}\label{alg:vcsetup}
\end{algfloat}

\begin{algfloat}[H]
{\fontsize{8}{11}\selectfont
 \begin{algorithmic}[1]
  \STATE Set $PK_F = \PP$
  \STATE Set $L_{\textrm{F}} = \epsilon$ (\ie\ an empty list is created).
 \end{algorithmic}
 }
 \caption{\vcfninit}\label{alg:vcfninit}
\end{algfloat}

\begin{algfloat}[H]
{\fontsize{8}{11}\selectfont
 \begin{algorithmic}[1]
  \STATE $(SK_{\textrm{Sig}}, VK_{\textrm{Sig}}) \leftarrow \Sig.\keygen(1^\lambda)$
  \STATE $SK_S = SK_{\textrm{Sig}}$
  \STATE $L_{\textrm{Reg}} =L_{\textrm{Reg}} \cup (S, VK_{\textrm{Sig}})$
 \end{algorithmic}
 }
 \caption{\vcregister}\label{alg:vcregister}
\end{algfloat}

\begin{algfloat}[H]
{\fontsize{8}{11}\selectfont
 \begin{algorithmic}[1]
  \STATE $t \leftarrow \tau$
  \STATE $\skabe{0} \leftarrow \ABE.\keygen(S, F, \mskabe{0},\mpkabe{0})$
  \STATE $\skabe{1} \leftarrow  \ABE.\keygen(S, \overbar{F},\mskabe{1}, \mpkabe{1}) $
  \STATE $UK_{L_F, t}^0  \leftarrow \ABEkeyupdate(L_{\textrm{F}}, t, \mskabe{0}, \mpkabe{0})$
  \STATE $UK_{L_F, t}^1  \leftarrow \ABEkeyupdate(L_{\textrm{F}}, t, \mskabe{1}, \mpkabe{1})$
  \STATE Output:  $EK_{\textrm{F,S}} = (\skabe{0}, \skabe{1}, UK_{L_F,t}^0, UK_{L_F,t}^1)$ and
                 		  $L_{\textrm{F}} = L_{\textrm{F}} \cup S$
 \end{algorithmic}
 }
 \caption{\vccertify}\label{alg:vccertify}
\end{algfloat}

\end{minipage}

\begin{minipage}{\textwidth}

\begin{algfloat}[H]
{\fontsize{8}{11}\selectfont
 \begin{algorithmic}[1]
 \STATE $t \leftarrow \tau$
  \STATE $(m_0, m_1) \rand \messagespace \times \messagespace$
  \STATE $b \rand \{0,1\}$
  \STATE $c_b \leftarrow \ABE.\enc(t, x, m_b, \mpkabe{0})$
  \STATE $c_{1-b} \leftarrow \ABE.\enc(t, x, m_{1-b}, \mpkabe{1})$
  \STATE Output: $\sigma_x = ({c_b}, {c_{1-b}})$ and 
                  		$VK_{F,x} = (g(m_b), g(m_{1-b}), L_{\textrm{Reg}})$
 \end{algorithmic}
 }
 \caption{\vcprobgen}\label{alg:vcprobgen}
\end{algfloat}

\begin{algfloat}[H]
 {\fontsize{8}{11}\selectfont
 \begin{algorithmic}[1]
  \STATE Input: $EK_{\rm F,S}=(\skabe{0}, \skabe{1}, UK_{L_F, t}^0, UK_{L_F,t}^1)$ and $\sigma_x=({c_b}, {c_{1-b}})$
  \STATE Parse $\sigma_x$ as $(c, c^\prime)$
  \STATE $d_0 \leftarrow \ABE.\dec(c, \skabe{0}, \mpkabe{0}, UK_{L_F, t}^0)$
  \STATE $d_1 \leftarrow \ABE.\dec(c^\prime, \skabe{1}, \mpkabe{1}, UK_{L_F, t}^1)$
  \IF{$d_0 \neq \perp \OR\ d_1 \neq \perp$}
  	\STATE b=0
  \ELSE
  	\STATE b=1
  	 \STATE $d_0 \leftarrow \ABE.\dec(c, \skabe{1}, \mpkabe{1}, UK_{L_F, t}^1)$
 	 \STATE $d_1 \leftarrow \ABE.\dec(c^\prime, \skabe{0}, \mpkabe{0}, UK_{L_F, t}^0)$
  \ENDIF
  \STATE $\gamma \leftarrow \Sig.\sign((d_b, d_{1-b}, S), SK_S)$
  \STATE Output: $\sigma_y = (d_b, d_{1-b}, S, \gamma)$
 \end{algorithmic}
 }
 \caption{\vccompute}\label{alg:vccompute}
\end{algfloat}

\begin{algfloat}[H]
 {\fontsize{8}{11}\selectfont
 \begin{algorithmic}[1]
  \STATE Input: $VK_{F,x}=(g(m_b), g(m_{1-b}), L_{\textrm{Reg}})$ and $\sigma_y=(d_b, d_{1-b}, S, \gamma)$
  \IF{$S \in L_F$ \AND $(S, VK_{\textrm{Sig}}) \in L_{\textrm{Reg}}$}
  	\IF{$\Sig.\verify((d_b,d_{1-b}, S),\gamma, VK_{\textrm{Sig}}) \rightarrow \accept$}
  		\IF{$g(m_b) = g(d_b)$}
  			\STATE Output $(\mu = d_b, \tau_{\sigma_y} = (\accept, S))$ %
  		\ELSIF{$g(m_{1-b}) = g(d_{1-b})$}
  			\STATE Output $(\mu = d_{1-b}, \tau_{\sigma_y} = (\accept, S))$   %
		 \ELSE
		 	\STATE Output $(\mu = \perp, \tau_{\sigma_y} = (\reject, S))$
		\ENDIF
  	\ENDIF
  \ENDIF
  \STATE Output $(\mu = \perp, \tau_{\sigma_y} = (\reject, \perp))$  %
 \end{algorithmic}
 }
 \caption{\vcblindverify}\label{alg:vcblindverify}
\end{algfloat}

\begin{algfloat}[H]
 {\fontsize{8}{11}\selectfont
 \begin{algorithmic}[1]
  \STATE Input: $VK_{F,x}=(g(m_b), g(m_{1-b}), L_{\textrm{Reg}})$, $\sigma_y=(d_b, d_{1-b}, S, \gamma)$, $b$, and $(\mu, \tau_{\sigma_y})$ where
  $\mu \in \{d_b, d_{1-b}, \perp\}$
  \IF{$\tau_{\sigma_y} = (\accept, S)$ \AND $g(\mu)=g(m_0)$}
  	\STATE Output $\tilde{y}=1$
  \ELSIF{$\tau_{\sigma_y} = (\accept, S)$ \AND $g(\mu)=g(m_1)$}
  	\STATE Output $\tilde{y}=0$
  \ELSE
  	\STATE 	Output $\tilde{y}=\perp$	
  \ENDIF
 \end{algorithmic}
 }
 \caption{\vcretrieve}\label{alg:vcretrieve}
\end{algfloat}

\begin{algfloat}[H]
{\fontsize{8}{11}\selectfont
 \begin{algorithmic}[1]
  \IF{$\tau_{\sigma_y} = (\reject,S)$}
  		\STATE $L_{\textrm{F}} = L_{\textrm{F}}\setminus S$
		\STATE Refresh\footnote{By refresh we mean, for example, increment $\tau$ if it is a counter} $\tau$
  		\STATE $t \leftarrow \tau$
  		\STATE $UK_{L_F, t}^0 \leftarrow \ABE.\keyupdate(L_{\textrm{F}}, t, \mskabe{0},\mpkabe{0})$
  		\STATE $UK_{L_F, t}^1 \leftarrow \ABE.\keyupdate(L_{\textrm{F}}, t, \mskabe{1},\mpkabe{1})$
  \FOR{ all $S^\prime \in L_{\textrm{F}}, S^\prime \neq S$}
  	\STATE Parse  $EK_{F,S^\prime}$ as $(\skabe{0}, \skabe{1}, UK_{L_F, t-1}^0, UK_{L_F,t-1}^1)$
          \STATE  Update and send $EK_{F, S^\prime} =   (\skabe{0}, \skabe{1}, UK_{L_F, t}^0, UK_{L_F,t}^1)$.
          \ENDFOR
    \ELSE
    	\STATE output $\perp$
    \ENDIF
 \end{algorithmic}
 }
 \caption{\vcrevoke}\label{alg:vcrevoke}
\end{algfloat}
\end{minipage}
}
\FloatBarrier
\ifnum\lncs=0
\subsection{Proof of Security}

We now give a theorem and proof that the construction presented above is secure against the games presented in Section~\ref{sect:security}.

\begin{theorem}\label{Thm}\sloppy
Given a secure revocable KP-ABE scheme in the sense of \INDSHRAtext\ (\INDSHRA)~\textnormal{\cite{attr:attr09}} for a class of functions $\mathcal{F}$ closed under complement, a signature scheme secure against \EUFCMA\ and a pre-image resistant hash function $g$, let $\VCcal{}$ be the verifiable computation scheme defined in Algorithms~\ref{alg:vcsetup}--\ref{alg:vcrevoke}. Then
$\VCcal{}$ is secure against Public Verifiability, Revocation and Vindictive Servers.
\end{theorem}

Informally, the proof of Public Verifiability relies on the \INDCPA\ security of the
underlying revocable KP-ABE scheme and the pre-image resistance of the function $g$.
Revocation relies on the \INDSHRA\ security of the revocable KP-ABE scheme.
Finally, security against Vindictive Servers relies on the \EUFCMA\ security of
the signature scheme such that a vindictive server cannot return an incorrect
result with a forged signature claiming to be from an honest server (note that
chosen message attack is required since the vindictive client could act like a
client and submit computation requests to get a valid signature). 
\ifnum\lncs=0
	We now present a more formal proof for Theorem~\ref{Thm}.
	\input{Proofs}
\else
	The formal proofs of security for this construction are left to the full version of the
	paper.
\fi
\fi

%% file: Theorem.tex
\section{Construction}

We provide a full construction of \PVCCA\ using revocable KP-ABE in Appendix~\ref{sect:construction}.
Informally, the scheme operates in the following way. 
\begin{enumerate}
 \item  $\vcsetup$  establishes public parameters and a master secret key by calling the $\ABEsetup$ algorithm twice. In the manager model, we would require these algorithms to be run over the same set of random coins when choosing random exponents for attribute group elements such that the set of values for \U\ is the same in both cases, even if the associated semantic meaning differs. Thus, an adversary cannot recognize which set of parameters a given ciphertext belongs to, and hence cannot break the blind verification property.
  This algorithm also initializes a list of registered servers $L_{\it Reg}$ and a time source $\tau$\footnote{$\tau$ could be a counter that is maintained in the public parameters or a networked clock.}.\label{step:constrsetup}
 \item  $\vcfninit$  initializes a list of servers $L_F$ authorized to compute function $F$.
 \item  $\vcregister$  creates a public-private key pair by calling the signature $\keygen$ algorithm. This is run by the \caa\ (or the manager in the manager model) and updates $L_{\it Reg}$ to include $S$.
 \item  $\vccertify$  creates the key $EK_{F,S}$ that will be used by a server $S$ to compute $F$ by calling the $\ABEkeygen$ and $\ABEkeyupdate$ algorithms twice -- once with a ``policy'' for $F$ and once with the complement $\overbar{F}$.  The algorithm also updates $L_F$ to include $S$.\label{step:constrkey}
 \item  $\vcprobgen$  creates a problem instance $\sigma_x=(c_0,c_1)$ by encrypting two randomly chosen messages, and a verification key $VK_{F,x}$ by applying a pre-image resistant hash function $g$ to the messages. The ciphertexts and verification tokens are ordered randomly according to a bit $b$, such that the positioning of an element does not imply whether it relates to  $F$ or for $\overbar{F}$. \label{step:constrprob}
 \item  $\vccompute$  is run by a server $S$ and computes $F(x)$.  Given a problem instance $\sigma_x=(c_0,c_1)$ it returns $(m_0,\perp)$ if $F(x) = 1$ or $(\perp,m_1)$  if $F(x) = 0$, ordered  according to $b$ chosen in \vcprobgen, together with a digital signature computed over the output. The server can determine the value of $b$ based on the results of decryptions with the different ABE parameters, and order his output correspondingly.
 \item  $\vcverify$  either accepts the output $\sigma_y=(d_0,d_1)$ or rejects it.  This algorithm verifies the signature on the output and then confirms the output is correct by applying $g$ and comparing with $VK_{F,x}$.
 In \vcblindverify\ the verifier can compare pairwise between the components of $\sigma_y$ and $VK_{F,x}$ to determine correctness but as they are unaware of the value of $b$, they do not know the order of these elements and therefore do not learn whether the correct output corresponds to $F$ or $\overbar{F}$ being satisfied \ie\ if $F(x)=1$ or $0$ respectively.
 The verifier outputs an \accept\ or \reject\ token as well as the satisfying (non-$\perp$) output value $\mu \in \{d_b, d_{1-b}\}$.
 In \vcretrieve\ a verifier that has knowledge of $b$ can check whether the output from \blindverify\ matches $m_0$ or $m_1$.
 \item  $\vcrevoke$  is run by the \caa\ and redistributes fresh keys to all non-revoked servers.  This algorithm updates $L_F$ and updates $EK_{F,S}$ using the results of two calls to the $\ABEkeyupdate$ algorithm.
\end{enumerate}

\begin{theorem}\label{Thm}\sloppy
Given a secure revocable KP-ABE scheme in the sense of \INDSHRAtext\ (\INDSHRA)~\textnormal{\cite{attr:attr09}} for a class of functions $\mathcal{F}$ closed under complement, a signature scheme secure against \EUFCMA\ and a pre-image resistant hash function $g$, let $\VCcal{}$ be the verifiable computation scheme defined in Algorithms~\ref{alg:vcsetup}--\ref{alg:vcrevoke}. Then
$\VCcal{}$ is secure in the sense of Public Verifiability, Revocation, Vindictive Servers, Blind Verification and Vindictive Managers.
\end{theorem}

Informally, the proofs of Public Verifiability and against Vindictive Managers rely on the \INDCPA\ security of the
underlying revocable KP-ABE scheme and the pre-image resistance of the function $g$.
Revocation relies on the \INDSHRA\ security of the revocable KP-ABE scheme. 
Blind Verification relies on the indistinguishability of two random messages, and the pre-image resistance of $g$.
These proofs are left for the full version of the paper.
Now we present a proof sketch for the security against vindictive servers.

\ifnum\lncs=0
	We now present a more formal proof for Theorem~\ref{Thm}.
	\input{Proofs}
\else
\fi

\begin{proof}[Sketch]
Let \advvc\ be an adversary with non-negligible advantage against the Vindictive Servers game (Game \ref{game:vindS}). 
We construct an adversary \advsig\ with non-negligble advantage $\delta$ in the \EUFCMA\ signatures game using \advvc.
$\advsig$ interacts with the challenger $\chall$ in the \EUFCMA\ security game and acts as the challenger for $\advvc$ in the 
security game for Vindictive Servers for a function $F$. Here the idea is that \advsig\ can create a \VC\ instance and play the Vindictive Servers game with \advvc\ by executing Algorithms~\ref{alg:vcsetup}--\ref{alg:vcrevoke} himself.
\advsig\ will guess a server identity that he thinks the adversary will select to vindictively revoke. The signature signing key that would be generated during the \register\ algorithm for this server will be implicitly set to be the signing key in the \EUFCMA\ game and any \compute\ oracle queries for this identity will be forwarded to the challenger to compute. Then, assuming that \advsig\ guessed the correct server identity, \advvc\ will output a forged signature that \advsig\ may output as its guess in the \EUFCMA\ game. If \advsig\ guessed the challenge identity correctly (i.e. $\overbar{S} = \tilde{S}$) then 
\advsig\ succeeds with the same non-negligible advantage $\delta$ as \advvc.
Let $n = \left\vert \U_{\sf{id}}\right\vert$, then the probability that \advsig\ correctly guesses $\overbar{S} = \tilde{S}$ is $\frac{1}{n}$ and 
$Adv_{\advsig} \geq \frac{1}{n}Adv_{\advvc}
			\geq \frac{\delta}{n} 
			\geq negl(\lambda)$.
Thus we conclude that \advsig\ has a non-negligible advantage against the \EUFCMA\ game if \advvc\ has a non-negligible advantage in the Vindictive Servers game, but as we assume the signature scheme in our construction to be \EUFCMA\ secure, such an adversary may not exist.
\end{proof}

%% file: AppendixGames.tex
\section{Security Games}
\label{sect:appgames}

\hspace{-15pt}
\begin{minipage}{1.2\textwidth}
\begin{minipage}[t]{0.45\textwidth}
\vspace{0pt}
\begin{gamefloat}[H]
 { \fontsize{8}{10}\selectfont
\begin{algorithmic}[1]
        \STATE $(\PP, \MSK) \leftarrow \setup(1^\lambda)$;
		\STATE $(PK_F, L_{{F}}) \leftarrow \fninit(\PP, \MSK, F)$;
		\STATE $SK_{S} \leftarrow \register(\PP, \MSK, S)$;
		\STATE $(EK_{F,S}, L_{{F}}) \leftarrow \certify(\PP, \MSK, F, L_{{F}}, S)$;
\STATE $(t^\star,x^\star) \leftarrow \adv^{\oraclesign}(PK_F, L_F, \PP)$;
		\STATE $(\sigma_{x^\star}, VK_{F,x^\star}, b) \leftarrow \probgen((t^\star, x^\star), PK_{F})$;
		\STATE $\sigma_{y}  \leftarrow \compute(\sigma_{x^\star}, EK_{F,S}, SK_S)$;
		\STATE $(\mu, \tau_{\sigma_y})\leftarrow \adv^{\oraclesign}(\PP, \sigma_y, VK_{F,x^\star}, PK_F, L_F)$;
		\STATE $\tilde{y} \leftarrow \retrieve(\mu, \tau_{\sigma_y}, VK_{F,x^\star}, b))$;
		\IF{$(\tilde{y} \neq F(x^\star))$ \AND $(\tilde{y} \neq \perp)$}
		\RETURN{$1$}
		             \ELSE
		             \RETURN{$0$}
		             \ENDIF
\end{algorithmic}
}
\caption{\gametitle{\adv}{VindictiveM}{\VCpub, F, 1^\lambda}}\label{game:vindM}
\end{gamefloat}
\end{minipage}
\begin{minipage}[t]{0.417\textwidth}
\vspace{0pt}
\begin{gamefloat}[H]
 {\fontsize{8}{10}\selectfont
\begin{algorithmic}[1]
        \STATE $(\PP, \MSK) \leftarrow \setup(1^\lambda)$;
		\STATE $(PK_F, L_{{F}}) \leftarrow \fninit(\PP, \MSK, F)$;
        \STATE $SK_{S} \leftarrow \register(\PP, \MSK, S)$;
		\STATE $(EK_{F,S}, L_{{F}}) \leftarrow \certify(\PP, \MSK, F, L_{{F}}, S)$;
		\STATE $x \rand \dom(F)$;
       \STATE $t \rand \tau$;
             \STATE $(\sigma_{x}, VK_{F,x}, b) \leftarrow \probgen((t, x), PK_{F})$;
       \STATE $\sigma_{y}\leftarrow \compute(\sigma_{x}, EK_{F,S}, SK_S)$;
       \STATE $\hat{b} \leftarrow \adv^{\oraclesign}(\sigma_{y},VK_{F,x}, \PP, PK_F, L_F))$;
	\IF{$\hat{b} = F(x)$}
             \RETURN{$1$}
             \ELSE
             \RETURN{$0$}
             \ENDIF
\end{algorithmic}
}
\caption{\gametitle{\adv}{BVerif}{\VCpub, F, 1^\lambda}}\label{game:outpriv}
\end{gamefloat}
\end{minipage}
\end{minipage}

\vspace{-10pt}

\subsubsection{Vindictive Manager.}

In Game~\ref{game:vindM} we capture security against vindictive managers attempting to provide the client with an incorrect answer, as discussed in Section~\ref{sect:vindm}. This is a natural extension of the Public Verifiability notion (Game \ref{game:pv}) in the manager model.
The adversary, on line 5, chooses a challenge input value $x$, and the server computes an encoded output of $F(x)$.
The adversary is then provided the encoded output and verification key and must output an encoded output $\mu$ and an acceptance token.
The challenger runs $\retrieve$ on $\mu$ to get an output value $\tilde{y}$, and the adversary wins if the challenger accepts this output and $\tilde{y} \neq F(x)$.
We remark that manager model instantiations may vary depending on the level of trust given to the manager.
A completely trusted manager may simply return the result to a client, whilst a completely untrusted manager may have to provide the full output from the server and the client performs the full \verify\ step as well (in this case, security against vindictive managers will reduce to Public Verifiability since the manager would need to forge a full encoded output that passes a full verification step).
Here we consider a middle ground where the manager is semi-trusted but the clients would still like a final, efficient  check.

\subsubsection{Blind Verification.}
\vspace{-10pt}

With Game~\ref{game:outpriv}, we aim to show that a verifier that does not know the value of $b$ chosen in \probgen\ cannot learn the value of $F(x)$ given the encoded output. 
The challenger chooses an input value, $x$, at random from the domain of $F$ and a time period, and uses these to generate an encoded input.
He runs \compute\ on this input and gives the encoded output and the verification key to the adversary who must output a guess for the value of $F(x)$.
We require that $\adv$ does not make a query to the \certify\ oracle for the function $F$ else he may use trial decryptions to compute $b$, and may not submit $x$ to the \probgen\ oracle. Similarly, $\adv$ may not use the \compute\ oracle for $\sigma_x$, but we assume he does not get this value and by the \INDCPA\ property of the ABE scheme we use, he may not guess a valid ciphertext for $x$.

%% file: mainCA.bbl
\begin{thebibliography}{1}

\bibitem{attr:attr09}
N.~Attrapadung and H.~Imai.
\newblock Attribute-based encryption supporting direct/indirect revocation
  modes.
\newblock In M.~G. Parker, editor, {\em IMA Int. Conf.}, volume 5921 of {\em
  Lecture Notes in Computer Science}, pages 278--300. Springer, 2009.

\bibitem{cart:white14}
H.~Carter, C.~Lever, and P.~Traynor.
\newblock Whitewash: Outsourcing garbled circuit generation for mobile devices.
\newblock Cryptology ePrint Archive, Report 2014/224, 2014.
\newblock \url{http://eprint.iacr.org/}.

\bibitem{CKKC}
S.~G. Choi, J.~Katz, R.~Kumaresan, and C.~Cid.
\newblock Multi-client non-interactive verifiable computation.
\newblock In {\em TCC}, pages 499--518, 2013.

\bibitem{GGP}
R.~Gennaro, C.~Gentry, and B.~Parno.
\newblock Non-interactive verifiable computing: Outsourcing computation to
  untrusted workers.
\newblock In T.~Rabin, editor, {\em CRYPTO}, volume 6223 of {\em Lecture Notes
  in Computer Science}, pages 465--482. Springer, 2010.

\bibitem{DBLP:conf/stoc/Gentry09}
C.~Gentry.
\newblock Fully homomorphic encryption using ideal lattices.
\newblock In M.~Mitzenmacher, editor, {\em STOC}, pages 169--178. ACM, 2009.

\bibitem{gold:multi13}
S.~Goldwasser, V.~Goyal, A.~Jain, and A.~Sahai.
\newblock Multi-input functional encryption.
\newblock Cryptology ePrint Archive, Report 2013/727, 2013.
\newblock \url{http://eprint.iacr.org/}.

\bibitem{Waters}
V.~Goyal, O.~Pandey, A.~Sahai, and B.~Waters.
\newblock Attribute-based encryption for fine-grained access control of
  encrypted data.
\newblock {\em IACR Cryptology ePrint Archive}, 2006:309, 2006.

\bibitem{PRV}
B.~Parno, M.~Raykova, and V.~Vaikuntanathan.
\newblock How to delegate and verify in public: Verifiable computation from
  attribute-based encryption.
\newblock Cryptology ePrint Archive, Report 2011/597, 2011.

\bibitem{DBLP:conf/focs/Yao86}
A.~C.-C. Yao.
\newblock How to generate and exchange secrets (extended abstract).
\newblock In {\em FOCS}, pages 162--167. IEEE Computer Society, 1986.

\end{thebibliography}
